\documentclass[%
 reprint,
superscriptaddress,
 amsmath,amssymb,
 aps,
pre,
]{revtex4-2}

\usepackage{graphicx}
\usepackage{dcolumn}
\usepackage{bm}
\usepackage{todonotes}

\begin{document}

\preprint{APS/123-QED}

\title{Comparison of different approaches to single-molecule imaging of enhanced enzyme diffusion
}

\author{Mengqi Xu}
\affiliation{Department of Physics, Syracuse University,  Syracuse, NY 13244}
\author{W. Benjamin Rogers}
\affiliation{Martin A. Fisher School of Physics, Brandeis University, Waltham, MA 02453 USA}
\author{Wylie W. Ahmed}
\affiliation{Department of Physics, California State University, Fullerton, CA 92831}
\author{Jennifer L. Ross}
\affiliation{Department of Physics, Syracuse University,  Syracuse, NY 13244}

\date{\today}

\begin{abstract}
Enzymes have been shown to diffuse faster in the presence of their reactants. Recently, we revealed new insights into this process of enhanced diffusion using single-particle tracking (SPT) with total internal reflection fluorescence (TIRF) microscopy. We found that the mobility of individual enzymes was enhanced three fold in the presence of the substrate, and the motion remained Brownian. In this work, we compare different experimental designs, as well as different data analysis approaches, for studying single enzyme diffusion. We first tether enzymes directly on supported lipid bilayers (SLBs) to constrain the diffusion of enzymes to two dimensions. This experimental design recovers the 3-fold enhancement in enzyme diffusion in the presence of the substrate, as we observed before. We also simplify our system by replacing the bulky polymers used in the prior chamber design with a SLB-coated surface and glycerol. Using this newly-designed SLB/glycerol chamber, we compare two different analysis approaches for SPT: the mean-squared displacement (MSD) analysis and the jump-length analysis. We find that the MSD analysis requires high viscosity and large particles to accurately report the diffusion coefficient, while jump-length analysis depends less on the viscosity or size. Furthermore, the SLB-glycerol chamber fails to reproduce the enhanced diffusion of enzymes because glycerol inhibits enzyme activity. 
\end{abstract}

\pacs{Valid PACS appear here}
\maketitle

\section{Introduction}
Enzymes, a group of nanoscale active proteins, are molecular machines that catalyze chemical reactions. In biological systems, enzymes bind with their substrates specifically at the active site and convert them into product molecules with high efficiency. For motor protein enzymes, they can harness the chemical free energy released during the substrate turnover and convert it into kinetic energy to achieve mechanical motion \cite{Tyska2002, Tsao2014, Hancock2014}.  
Recently, this capability of chemical/kinetic energy conversion was also extended to many other types of active enzymes, such as urease, catalase, DNA polymerase, and hexokinase. These enzymes were shown to diffuse faster when catalyzing their substrates in the solution \cite{Borsch1998, Muddana2010, Riedel2015, Jee2018, Xu2019, Sengupta2013, Illien2017, Zhao2018, Jee2020}. The mechanism of the molecular level enhanced diffusion of single enzyme is still under debate. 

Most of the fundamental studies on the enhanced diffusion of enzymes exploited fluorescence correlation spectroscopy (FCS) for diffusion measurements \cite{Borsch1998, Muddana2010, Riedel2015, Jee2018, Sengupta2013, Illien2017, Zhao2018}. However, recently, Gunther et al. showed that typical FCS experiments might introduce artifacts in diffusion measurements for enzymes, calling some of the former findings into question \cite{Gunther2018}. In an attempt to test if the enhanced diffusion is real or just a result of experimental artifacts, researchers have employed a variety of alternative techniques to provide complementary measurements for enzyme diffusion \cite{Zhang2018, Gunther2019, Jee2019, Chen2020}. These new techniques refuted some prior reports of enhanced diffusion. For example, aldolase measured by dynamic light scattering (DLS) \cite{Zhang2018} or nuclear magnetic resonance (NMR) \cite{Gunther2019}, and alkaline phosphatase detected by anti-Brownian electrokinetic (ABEL) trap \cite{Chen2020}, showed no enhanced diffusion. 

In our recent paper \cite{Xu2019}, we verified the enhanced diffusion of urease in the presence of its substrate, urea, by using a direct single-molecule imaging method with total internal reflection fluorescence (TIRF) microscopy. We found that the overall mobility of each individual active urease was increased by 2--3 fold at saturated substrate concentration, while diffusion remained Brownian. 
Although we were able to recapitulate the enhanced diffusion of enzymes using SPT, there were caveats to this method: (1) We recorded the 2D projections of 3D trajectories. Specifically, the TIRF microscope exploits the total internal reflection of incident light to form an evanescent field immediately adjacent to the interface between the specimen and the glass coverslip. Thus, only fluorophores within the 200-nm excitation region above the surface are capable of being excited, which makes single-particle imaging possible. The motion in z-direction (perpendicular to the interface) is totally lost. (2) Polymers in solution and on the surface might have unquantifiable effects on enzyme diffusion. In our prior work, to slow down the enzyme mobility and facilitate tracking, we introduced a surface polymer coating (Pluronic F127) and a viscous polymer (methylcellulose) to our experimental chamber. The presence of these additives indeed slowed down the enzyme mobility enough for accurate tracking, but also raised concerns of their potential, unknown effects on enzyme diffusion that were not easily quantifiable. 


Here, we dissect the experimental caveats of our prior work, and compare different experimental designs as well as data analysis approaches of direct single-molecule imaging of enzyme diffusion measurements. We first tether urease directly to a supported lipid bilayer (SLB) to constrain the enzyme diffusion to two dimensions (Fig. \ref{fig:setup} A). We measure the diffusion of tethered urease with and without urea, and show a 3-fold enhancement in the diffusion of urease, as we observed before \cite{Xu2019}. To avoid adding polymers and to minimize the environmental complexity for enzyme diffusion, we replace the polymer brush coated surface with a SLB, and substitute the large viscous polymer, methylcellulose, with a smaller, well-characterized molecular viscosity agent, glycerol (Fig. \ref{fig:setup} B, C). We test the reliability of this new experimental design by measuring the diffusion of particles with various sizes ($R$ = 2.3, 4.8, 7.0, 99 nm) and in solutions of different viscosities ($\eta$ = 2.73, 6.86, 12.76, 26.85, 41.30, 66.65, 113.85, 208.13 mPa$\cdot$s). We find that the diffusion coefficients scale with particle size and solvent viscosity as expected from the Stokes-Einstein equation. We also compare two different data analysis approaches for SPT: the mean-squared displacement (MSD) analysis \cite{Tarantino2014} and the jump-length analysis \cite{Hansen2018}. We find that the MSD analysis requires high viscosity and large particle size to accurately report the diffusion coefficient, while the jump-length analysis depends less on the viscosity or size. Using this newly-designed SLB/glycerol chamber, we repeat the urease diffusion experiments in different concentrations of urea. However, this new chamber design fails to reproduce the enhanced diffusion of urease even in saturated urea concentration. We attribute this failure to the inhibited catalytic activity of urease due to the presence of glycerol.

\section{methods}

\subsection{Enzyme preparation and activity assay}

Experiments are prepared using commercially-available reagents. Urease from Jack Bean is purchased from TCI Chemicals. Aldolase from rabbit muscle is purchased from Sigma Aldrich. Green fluorescent protein (GFP) is purified following a standard protocol for His-tagged protein purification. Sub-micron multi-color plastic spheres ($R$ = 99 nm) are purchased from Thermo Fisher. Enzymes are fluorescently labeled with Alexa Fluor 647 $C_2$ maleimide (Thermo Fisher) using a commercially available protein labeling kit following the optimized protocols provided by Thermo Fisher. Biotinylated enzymes are made by using a commercially supplied EZ-Link Sulfo-NHS-LC-Biotinylation kit (Thermo Fisher) following the instructions provided. The urease activity assay is performed following a published protocol in Ref.\cite{Activity}. Briefly, we use phenol red as an color indicator which turns from yellow to red as pH increases, to estimate the urease activity. The assay mixture contains 10 nM urease, 28 $\mu$M phenol red, 2.5 mM urea, and 30\% or 75\% glycerol or 1$\times$ PBS buffer to contribute to a total volume of 1 ml. We measure the absorbance at 560 nm every 6 seconds to quantify the color-changing rate using UV-vis spectroscopy.

\begin{figure}
	\centering
		\includegraphics[width=1\linewidth]{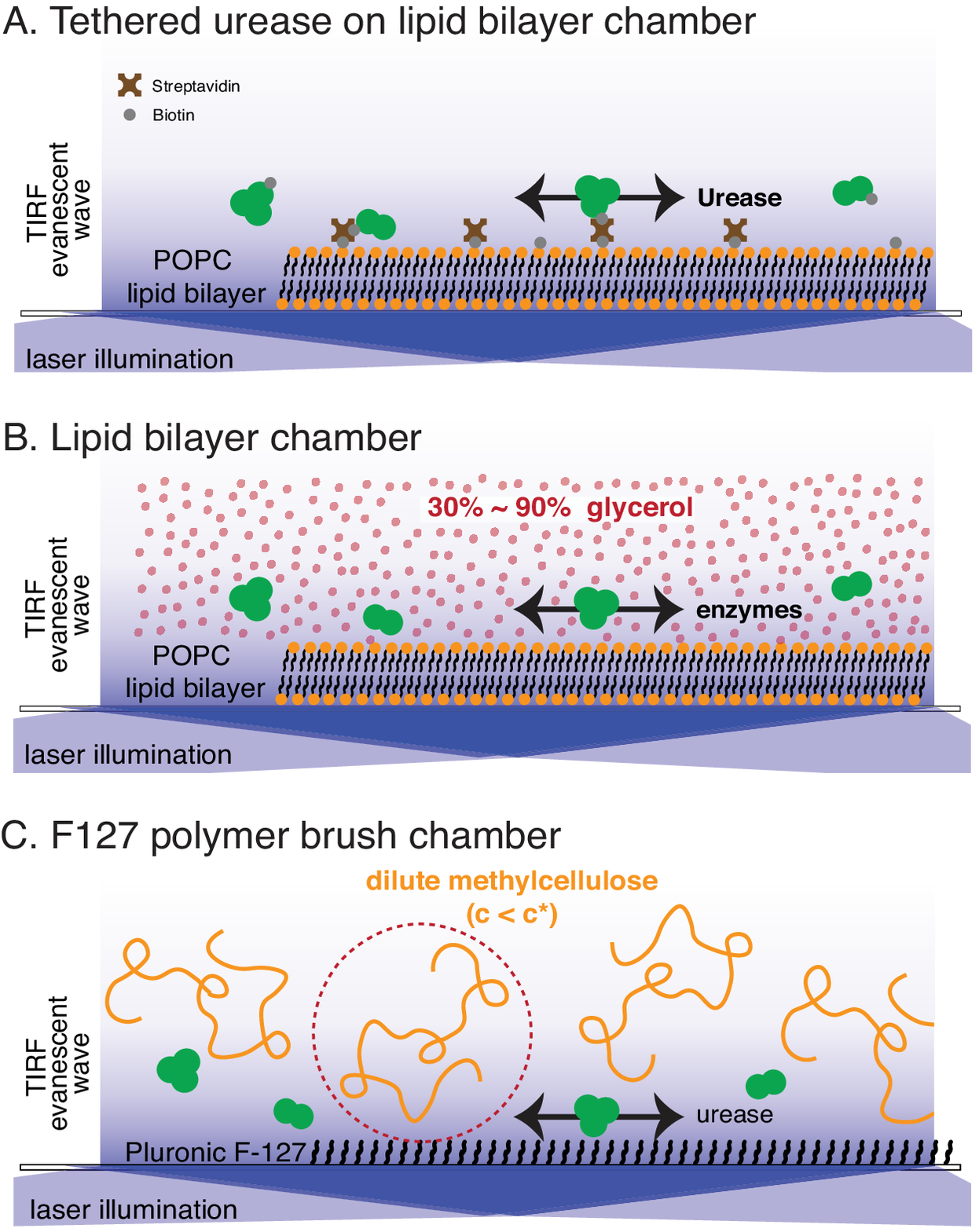}
		\caption{Schematics of experimental chamber designs. A) The tethered urease experimental design in which biotin modified urease is tethered on a biotinylated SLB via biotin(gray)-streptavidin(brown) interactions. B) The SLB/glycerol chamber, where the surface is coated with SLB (orange and black) and a certain percentage of glycerol (red) is added as a viscous agent to slow down the mobility of enzymes. C) The F127 polymer brush chamber design used in our prior work \cite{Xu2019}. Surface was coated by F127 block-copolymer (black); 3\% dilute methylcellulose polymers (orange, $R_g \sim$ 30 nm \cite{MC2015}) were used to slow down the mobility.}
	\label{fig:setup}
\end{figure}

\subsection{Chamber setup}

Experimental flow chambers are made from a glass slide, a cover slip (No. 1.5 Fisherbrand, Thermo Scientific), and two pieces of double stick tape. The tape is sandwiched between the slide and the cover slip, acting as a spacer and forming a 5-mm-wide channel. Thus, the chamber volume is limited to $\sim 10$ $\mu$l by the thickness of the tape ($80 \sim 100$ $\mu$m in height).



\subsubsection{Supported lipid bilayers}

The supported lipid bilayers are made by fusing small unilamellar vesicles (SUVs) on the chamber surface \cite{SLB}. SUVs are made of POPC (1-palmitoyl-2-oleoyl-glycero-3-phosphocholine) purchased from Avanti. For biotinylated lipid surfaces, another 0.5 mol\% biotin-PE (1,2-dioleoyl-sn-glycero-3-phosphoethanolamine-N-(cap biotinyl)) is added (Avanti). First, 40~$\mu$l of 10~mg/ml POPC in chloroform is dissolved in 70~$\mu$l of chloroform and mixed well. Chloroform is then evaporated from the mixture under a gentle stream of N$_2$ gas for 10 min. The lipid mixture is further dried out in a vacuum desiccator for 30 min. The dried lipid is rehydrated in 100 $\mu$l PBS buffer and vortexed for 1 min to form giant unilamellar vesicles (GUVs). The white opaque GUV suspension is then sonicated using a sonicator microtip probe (Sonifier) for 3 min to form a clear SUV solution. This clear SUV solution is stored at 4$^{\circ}$C and used for the SLB surface coating. 

To make SLB-coated flow chambers, 10 $\mu$l of SUV solution is first flowed in and incubated for 20 min to allow for the fusion of SUVs to the surface. Excess unfused SUVs are subsequently removed by washing the chamber with PBS buffer 7 times. The SLB-coated chambers are kept in a humid container to prevent dehydration and taken out immediately before use. A certain percentage of glycerol is also added to slow down the diffusion of enzymes in the SLB chamber (Fig. \ref{fig:setup} B).

\subsubsection{Tethered urease on SLB}

 Urease is first modified with biotin and fluorescently labelled with Alexa647 using commercially available kits following the protocols provided. Then 5.38 $\mu$M Alexa647-biotin-urease is mixed with a 2-fold molar excess of streptavidin and incubated on ice for 1 hr to form SA-Alexa647-biotin-urease complexes. The reaction mixture is then diluted by 10,000 times to make the enzyme concentration optimized for single particle imaging. Biotinylated SLB-coated flow chambers are made following the same procedures as described above using biotinylated SUVs. Finally, 14 $\mu$l of diluted reaction mixture is flowed into the biotinylated SLB-coated chamber and incubated in a humid container for 10 min. Free unattached streptavidin and enzyme complex are then removed by washing with PBS buffer for 7 times (Fig. \ref{fig:setup} A).\\

For all chambers, an oxygen scavenging system (10 mM dithiothreitol (DTT), 15 mg/ml glucose, 0.15 mg/ml catalase, and 0.05 mg/ml glucose oxidase) is added to extend the lifetime of the fluorescent dyes and minimize photobleaching. For polymer brush and lipid surface chambers, the oxygen scavenging system is added to the chamber with the enzymes (diluted in PBS $\sim$ 100 pM) and the urea. For tethered enzyme chambers, the oxygen scavengers are added directly prior to imaging with the urea. All chambers are imaged using a custom-built TIRF microscope immediately after loading and kept measuring for a maximum of 30 minutes before discarding. 

\subsection{TIRF imaging }

Single-particle imaging is performed using total internal reflection fluorescence (TIRF) microscopy with a custom-built laser system (50 mM 488 nm laser and 100 mW 638 nm laser from CrystaLaser) constructed around a Nikon Ti-E microscope. Imaging is performed with a $60 \times$, 1.49 NA TIRF objective (Nikon), and then magnified an additional $2.5 \times$ before being projected onto an EM-CCD camera (IXON electron-multiplier CCD, Andor). The camera has $512 \times 512$ square pixels of 16.2 $\mu$m on each side, giving a final magnified pixel size of 107 nm/pixel. Movies were recorded at a rate of 17 frames/s ($\Delta t = 60$ ms/frame, ROI = 512$\times$512 pixels) or 105 frames/s ($\Delta t = 9.5$ ms/frame, ROI = 512$\times$76 pixels) with a 30 ms or 4 ms exposure using the Nikon Elements software. Laser power and EMCCD gain settings were kept constant for all movies.

\subsection{Data analysis}

\subsubsection{MSD analysis}

The mean squared displacement (MSD) analysis is performed using the same protocol as described in Ref. \cite{Xu2019}. A popular tracking plugin in ImageJ/FIJI, called ParticleTracker 2D/3D \cite{Sbalzarini2005}, is used to extract trajectories from microscopy videos. Homemade MATLAB codes based on Ref. \cite{Tarantino2014} are applied for trajectory analysis. For each trajectory, we compute the time-averaged mean squared displacements (MSD) over different lag times by $\big<(\Delta r_i(t))^2\big>=\big<[\vec{r}_i(\tau+t)-\vec{r}_i(\tau)]^{2}\big>_{\tau}$, where $r_i (t)$ is the position of the $i$th trajectory at lag time $t$, and the brackets $\big<\cdot\cdot\cdot\big>$ indicate a time average over $\tau$. We plot the MSDs as a function of lag time $t$, and derive the diffusion coefficient, $D$, from the slope of the MSD plot after fitting to the Einstein's diffusion equation in 2D:
    \begin{equation}
    \big<(\Delta r_i)^2\big>=4Dt
    \end{equation}
Since the diffusion coefficients obtained from SPT measurements follow a log-normal distribution empirically \cite{Saxton1997, Rossier2012, Bo2013, Knight2015}, we first log-transform the diffusion coefficients extracted for each experimental group and bin it into a histogram. Each histogram is then fit with a Gaussian, for which the mean is taken as the apparent diffusion constant after transforming back to the normal $D$ scale. 

The parameters used in ParticleTracker 2D/3D plugin are: Particle size = 3-5 pixels; cutoff = 0.001; Percentile = 1\%-5\%; Link range = 4; Displacement = 4-7; Dynamics type = Brownian, for optimal tracking. Usually, thousands of trajectories can be detected for each experimental group by ParticleTracker 2D/3D, but not all are used for MSD analysis. Two thresholds are applied to select trajectories for analysis: 1) the minimum trajectory length, $N$, and 2) the goodness of the MSD-fit, $R^2$. In our analysis, only trajectories of at least 10 frames ($N \geq$ 10) with the goodness of MSD-fit greater than 0.9 ($R^2 \geq$ 0.9) contribute to the histogram of logarithmic diffusion coefficients for each case.






\subsubsection{Jump-length analysis}

Trajectories are also analyzed by the ``jump-length'' method, which uses the statistics of jump-lengths (the displacements of particles over different lag times) to deduce the corresponding diffusion properties \cite{Hansen2018}. Briefly, for a particle starting at the origin and freely diffusing in 2D, the probability of finding it at position $r$ after a lag time $\Delta t$ can be described by: 
    \begin{equation}
    \label{eqn:jump-length}
    P(r, \Delta t)=\frac{r}{2D\Delta t}\exp\big[\frac{-r^2}{4D\Delta t}\big]    
    \end{equation}
where $D$ is the diffusion coefficient. Thus, by fitting the jump-length distributions of particles for different lag times to the above probability function, the diffusion coefficient can be assessed. We use a semi-analytical kinetic model-based jump length analysis called Spot-On to perform the jump-length analysis. This model was developed by Hansen et al in Ref.\cite{Hansen2018}. In their model, several factors are taken into account to distinguish different diffusion ensembles from the population and to compensate the biases from fast-moving particles, such as 'motion-blur'. In this approach, all trajectories detected by the tracking algorithm are treated equally and contribute to the jump length histogram with no thresholds applied. This model-based jump-length analysis approach was first adopted in single enzyme diffusion experiments by Chen et al. in Ref.\cite{Chen2020}. In their work, single enzymes were localized and tracked using a custom-written MATLAB implementation of the multiple-target tracing algorithm (MTT algorithm) \cite{Serge2008}. To be consistent, we exploit the same MTT algorithm for particle tracking in the following jump-length analysis. 

We use the following settings for the MTT algorithm: LocalizationError = -6.25; EmissionWavelength = 647; ExposureTime = 60 or 9.5 ms; NumDeflationLoops = 0; MaxExpectedD ($D_\textrm{max}$) = 1.4-70 $\mu$m$^2$/s;  NumGapsAllowed = 3 or 2 (see Supplemental Information for more details about the parameter settings in MTT). For Spot-On analysis we use: TimeGap = 60 or 9.5 ms; GapsAllowed = 2 or 1; dZ = 0.700; TimePoints = 2; UseEntireTraj = yes; D\_Free\_2State = [0.01, $D_\textrm{max}$ used in MTT]; D\_Bound\_2State = [0.0001, 0.001]; ModelFit = CDF. 

During the analysis, we find that the diffusion coefficient reported by Spot-On analysis depends linearly on the $D_\textrm{max}$ that we set in the MTT algorithm. To determine the optimized $D_\textrm{max}$ for each case, we repeat the analysis over different $D_\textrm{max}$ values and find the corresponding $D$ from Spot-On. We plot the $D$ as a function of $D_\textrm{max}$, and expect three regimes: 1) a linear regime in which $D$ increase as $D_\textrm{max}$ increases; 2) a plateau when $D$ no longer depends on $D_\textrm{max}$; 3) another linear regime when $D$ is proportional to $D_\textrm{max}$ again. We use the $D_\textrm{max}$ value at the plateau as the most optimized value for each case.
 

\section{Results}

\subsection{Enhanced Diffusion of Urease Tethered to SLB}

\begin{figure}[]
	\centering
		\includegraphics[width=1\linewidth]{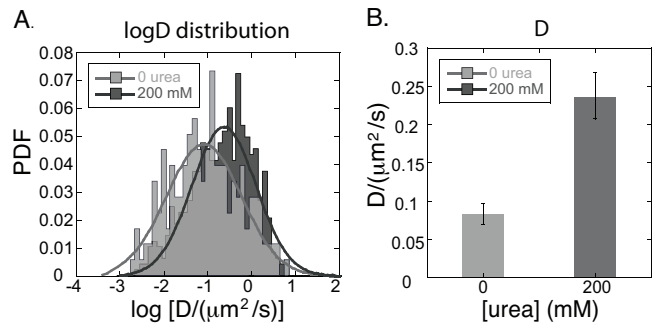}
		\caption{A) Histograms of logarithmic diffusion constant log$D$ of tethered urease with (dark gray, $N$ = 484) and without (gray, $N$ = 178) the presence of 200 mM urea. Line: the corresponding Gaussian fit to each log$D$ histogram. B) Apparent diffusion coefficients derived from the mean of the Gaussian fits for tethered urease with (dark gray) and without (gray) the presence of 200 mM urea. Error bars are determined from the standard errors of the mean of the Gaussian fits. All fit parameters are given in the supplemental information.}
	\label{fig:tethered urease}
\end{figure}

To confine the enzymes to 2D for accurate tracking, we tether the urease to the SLB surfaces. Each fluorescently labeled streptavidin-urease (SA-urease) complex is bound to the biotinylated SLB via biotin-streptavidin interactions and imaged under TIRF microscopy (Fig. \ref{fig:setup} A). Since the SLB is fluid, tethered urease could still diffuse freely in 2D. We measure the mobility of each tethered urease with and without the presence of urea. Fig. \ref{fig:tethered urease} A shows the distributions of log-transformed diffusion coefficient of tethered urease in the absence or presence of 200 mM urea. Each log$D$ histogram is fit by a Gaussian. The mean of each Gaussian fit is transformed back to normal diffusion units and used as the diffusion coefficient for each case (Fig. \ref{fig:tethered urease} B). For the buffer case, we find $D_\textrm{buffer}=0.0824$ $\mu$m$^2$/s, and for the urea group we have $D_\textrm{urea}=0.236$ $\mu$m$^2$/s, an almost 3-fold enhancement in diffusion. This result is an important quantitative confirmation of our prior result demonstrating enhanced diffusion \cite{Xu2019} using an independent experimental approach to perform the SPT and MSD analysis. 


\subsection{Single Molecule Imaging Using SLB/glycerol Chamber}

In our prior work, we used a surface polymer coating (Pluronic F127) and a viscous polymer (methylcellulose) to slow down the enzyme mobility and facilitate tracking. However, these bulky polymers also added complexity to the environment of the enzymes diffusion. These additives were held constant between experiments, but still could cause effects that are hard to quantify. To minimize the chamber complexity, we replace the two large, complex polymers with a well-characterized SLB surface, and a small, molecular viscosity agent, glycerol. We first measure the diffusion of different particles in solutions of different viscosities to test the reliability of this new experimental design. Under the same scenario, two data analysis approaches, MSD analysis and the jump-length analysis, are compared. Using this newly-designed SLB/glycerol chamber, we then examine the diffusion of urease in different concentrations of urea under two different viscosities.

\subsubsection{Comparison of MSD Analysis And Jump-length Analysis }

\begin{figure}
	\centering
		\includegraphics[width=1\linewidth]{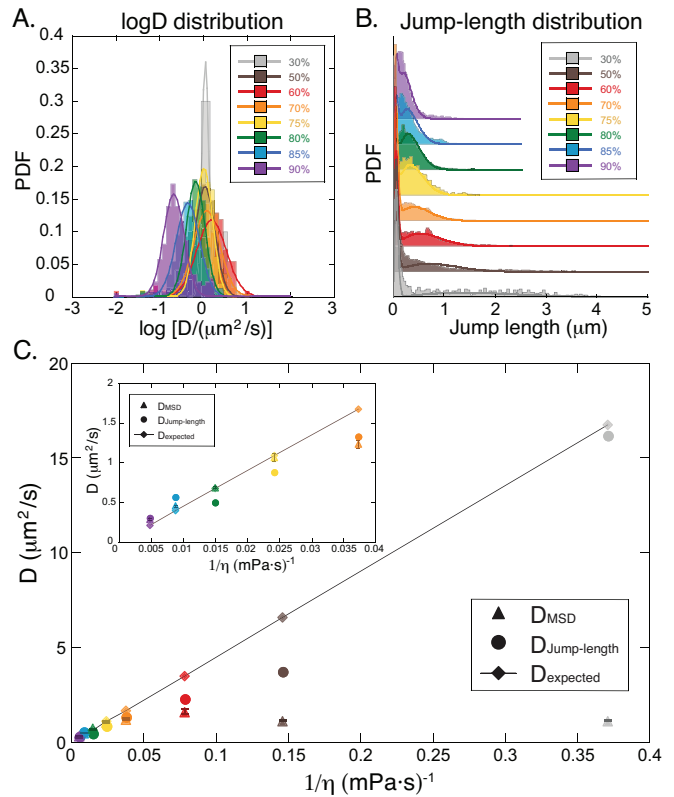}
		\caption{ A) Histograms of logarithm of diffusion coefficients for aldolase under different glycerol percentages: 30\% (gray region, $N$ = 10), 50\% glycerol (brown region, $N$ = 420), 60\% (red region, $N$ = 109), 70\% (orange region, $N$ = 313), 75\% (yellow region, $N$ = 97), 80\% (green region, $N$ = 676), 85\% (blue region, $N$ = 736), 90\% (purple region, $N$ = 213). Colored lines show the corresponding Gaussian fits to the log$D$ histograms. B) Distributions of jump-length for aldolase over one lag time $\Delta t = 60$ ms at different glycerol percentages. Colored lines are the corresponding kinetic model fits from Spot-On analysis. C) Comparisons of $D_\textrm{MSD}$ (triangle), $D_\textrm{jump-length}$ (circle) and $D_\textrm{expected}$ (diamond) of aldolase at different viscosities. Inset: enlarged version of the high viscosity regime with glycerol\% $\geq$ 70\%. Guide line shows the linear relationship between $D$ and 1/$\eta$ suggested by the Stokes-Einstein equation $D=k_BT/6\pi\eta R$. Error bars are determined from the standard errors of the mean of the Gaussian fits. All fit parameters are given in the supplemental information.}
	\label{fig:aldolase viscosity}
\end{figure}

Freely diffusing nanoscale particles usually undergo Brownian motion. Their diffusion coefficients are described by the Stokes-Einstein equation: $D=\frac{k_BT}{6\pi\eta R}$. Normally, the diffusion constant $D$ scales inversely with solvent viscosity $\eta$ and particle size $R$. To test the efficacy of the SLB/glycerol chamber, we first fix the particle size and measure how the diffusion coefficient changes with viscosity. We examine the diffusion of a single type of enzyme, aldolase ($R$ = 4.8 nm \cite{Raldolase}), for different viscosities. Solvent viscosity is tuned by varying the percent volume of glycerol in the buffer solution. We applied two different data analysis approaches, MSD analysis and the jump-length analysis, to analyze the same set of enzyme diffusion videos. Analysis results from each method are then compared side-by-side. 

We first performed the MSD analysis following the same protocols as described in our prior work \cite{Xu2019}. The distributions and Gaussian fits of log-transformed diffusion coefficients of aldolase under different glycerol percentages are shown in Fig. \ref{fig:aldolase viscosity} A. The mean log$D$ of each distribution, after transforming back to the typical diffusion units, is used as the apparent diffusion coefficient $D_\textrm{MSD}$ for each case (triangle, Fig. \ref{fig:aldolase viscosity} C). We also applied the jump-length analysis (Spot-On) on the same data set. We find the distributions of aldolase jump lengths for each glycerol percentage (Fig. \ref{fig:aldolase viscosity} B), and derive the diffusion coefficient $D_\textrm{jump-length}$ from the kinetic model fitting following the procedures as described in Ref. \cite{Chen2020} and Methods (circle, Fig. \ref{fig:aldolase viscosity} C ). We compare the diffusion coefficients derived from MSD analysis, $D_\textrm{MSD}$, jump-length analysis, $D_\textrm{jump-length}$, and their expected value determined by the Stokes-Einstein equation, $D_\textrm{expected}$, in Fig. \ref{fig:aldolase viscosity} C. In order to highlight the linear dependence of diffusion on the inverse of the viscosity, $D_\textrm{MSD}$, $D_\textrm{jump-length}$ and $D_\textrm{expected}$ are plotted as a function of the inverse of the viscosity $1/\eta$. 

We find that, at high viscosity ($1/\eta < 0.025$ (mPa$\cdot$s$)^{-1}$, or glycerol$\%>70\%$), both analysis methods report $D$ similar to $D_\textrm{expected}$ (Fig. \ref{fig:aldolase viscosity} C, inset). $D_\textrm{MSD}$ seems to match with the expected value better than $D_\textrm{jump-length}$ at high viscosities (see supplemental information for more detailed data). However, at low viscosity ($1/\eta > 0.08$ (mPa$\cdot$s$)^{-1}$, or glycerol$\%<60\%$), $D_\textrm{MSD}$ shows to be lower than to the expected value, deviating from the linear dependence on $1/\eta$ and appearing to plateau as viscosity decreases (Fig. \ref{fig:aldolase viscosity} C, triangle). We attribute this underestimation of the diffusion constant to the under-counting of fast-diffusing particles when using MSD analysis. Specifically, at low viscosity, particles move faster and exit the focal plane quickly, making it hard to track and acquire long trajectories. Only slowly moving particles, like large protein aggregates, remain in focus for long enough to be captured. Since our MSD analysis only analyzes trajectories of at least $N$ frames (usually, $N=10$ in our experiments), the short trajectories from fast-diffusing molecules are completely filtered out during the analysis process. As a result, slow-diffusing tracks account for the vast majority of trajectories analyzed, leading to an underestimation of the overall diffusion coefficient reported. From the data, we suggest that a minimum viscosity, $\eta_c=26.8$ mPa$\cdot$s (or glycerol$\%=70\%$), is required for MSD analysis to yield valid diffusion coefficients. While MSD analysis fails at low viscosities, we find that the jump-length analysis results are still close to $D_\textrm{expected}$ (Fig. \ref{fig:aldolase viscosity} C, circle). Thus, jump-length analysis seems more appropriate for the low viscosity regimes than the MSD analysis. 

We notice that the diffusion coefficients given by jump-length analysis seem to depend heavily on one of the parameters set in the tracking algorithm, called $D_\textrm{max}$. This parameter defines an area that a particle is assumed to explore during one lag time. This area is then used to search for the same particles between frames to connect and form trajectories. When using larger $D_\textrm{max}$ values, the possibility to mistakenly link two different particles into the same trajectory is increased, which in turn could lead to the report of a higher diffusion coefficient than expected. Therefore, care must be taken when choosing parameters for jump-length analysis.

\begin{figure}
	\centering
		\includegraphics[width=1\linewidth]{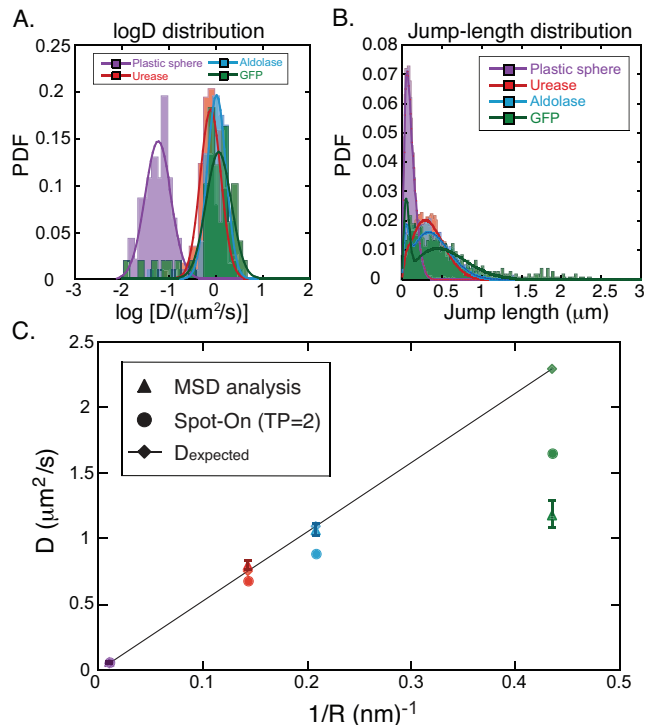}
		\caption{ A) Histograms of log-transformed diffusion coefficients for different-size particles in 75\% glycerol: plastic spheres (purple region, $N$ = 46), urease (red region, $N$ = 113), aldolase (blue region, $N$ = 97), GFP (green region, $N$ = 49). Colored lines show the corresponding Gaussian fits to log$D$ histograms. B) Distributions of jump-length at one lag time $\Delta t = 60$ ms for different-size particles in 75\% glycerol. Colored lines are the corresponding kinetic model fits from Spot-On analysis for the jump-length distributions. C) Comparisons of $D_\textrm{MSD}$ (triangle), $D_\textrm{jump-length}$ (circle) and $D_\textrm{expected}$ (diamond) for different-size particles in 75\% glycerol. Guide line represents the inverse proportional relationship between $D$ and $R$ suggested by the Stokes-Einstein equation $D=k_BT/6\pi\eta R$. Error bars are determined from the standard errors of the mean of the Gaussian fits. All fit parameters are given in the supplemental information.}
	\label{fig:size scaling}
\end{figure}

We next examine how well the SLB/glycerol chamber could work for measuring the diffusion of different-size particles at a fixed viscosity ($\eta=41.3$ mPa$\cdot$s, or glycerol$\%=75\%$). Similarly as before, we adopt two approaches, the MSD analysis and the jump-length analysis, to analyze the diffusion videos. We quantify the diffusion of four species of particles:  GFP ($R=2.3$ nm \cite{RGFP}), aldolase ($R=4.8$ nm \cite{Raldolase}), urease ($R=7.0$ nm \cite{Rurease}), and sub-micron multi-color plastic spheres ($R=99$ nm). From MSD analysis, we plot histograms of logarithmic diffusion coefficient for the four different particles (Fig. \ref{fig:size scaling} A). The apparent diffusion coefficient for each particle species is determined by the mean of each Gaussian fit after transforming log$D$ back to the normal $D$ scale as described before (triangle, Fig. \ref{fig:size scaling} C). For jump-length analysis, the distributions of jump-length over one lag time are shown for each particle species (Fig. \ref{fig:size scaling} B). The corresponding diffusion coefficients are derived from the kinetic model fitting in Spot-On and depicted as circles in Fig. \ref{fig:size scaling} C. Again, we plot $D$ as a function of the inverse radius $1/R$ to demonstrate the inversely proportional relationship between diffusion rate $D$ and particle size $R$ more clearly. 

We find that for relatively large particles, including the plastic sphere, urease, and aldolase, both MSD analysis and jump-length analysis report diffusion coefficients that match with the expected values (Fig. \ref{fig:size scaling} C). For smaller particles, such as GFP, MSD analysis again underestimates the expected diffusion coefficients. This underestimation of diffusion coefficients of smaller particles is likely due to the same issue described previously for low viscosity. Since smaller particles diffuse faster, most of the short trajectories from small fast-moving particles would be filtered out, leading to an under-counting of the fast population, which in turn results in the slower diffusion reported. As above, this suggests that a minimum size threshold, $R_c$, should be set for MSD analysis when using SLB/glycerol chambers to get reasonable diffusion measurements. Jump-length analysis seems to report a diffusion coefficient closer to the expected value, implying an advantage in analyzing fast-diffusing particles.  

In conclusion, in an attempt to perform better SPT experiments, we replace the bulky polymers with well-characterized molecular components: lipids on the surface and glycerol in solution. We examine the the reliability of this new SLB/glycerol chamber by measuring the diffusion of different-size particles in different viscosity solutions. We find that the particle diffusion in SLB/glycerol chamber behaves as the Stokes-Einstein equation suggests. The measured diffusion constants scale inversely with solvent viscosity $\eta$ and particle size $r$, which confirms the efficacy of SLB/glycerol chamber for SPT experiments. We also compare two data analysis methods: the MSD analysis and the jump-length analysis, using the SLB/glycerol chamber. We find that the MSD analysis is reliable at high viscosity and large particle size, which is the physical situation needed for relatively slow diffusion. Jump-length analysis seems to have less limitations on solvent viscosity and particle size, and has the advantage that it can analyze fast-moving particles. However, the diffusion coefficients reported from jump-length analysis depend strongly on the tracking parameters settings, specifically, $D_\textrm{max}$. Thus, care must be taken when choosing parameters for this approach. Also, all key parameters used for the data analysis should be reported to ensure the reproducibility of the results, as suggested in Ref. \cite{Hansen2018}.

\subsubsection{Enhanced Diffusion of Active Urease Using SLB/glycerol Chamber}

We next seek to reproduce the enhanced diffusion of active urease with the presence of urea using the newly-designed SLB/glycerol chamber. We find in the former section that MSD analysis performs better at high viscosity regimes, while jump-length analysis is more preferable at low viscosity environments. Thus, we make two sets of diffusion measurements on urease in two viscosity regimes: 1) the high viscosity regime using 75\% glycerol analyzed by MSD analysis and 2) the low viscosity regime using 30\% or no glycerol analyzed by jump-length analysis.\\

\begin{figure}
	\centering
		\includegraphics[width=1\linewidth]{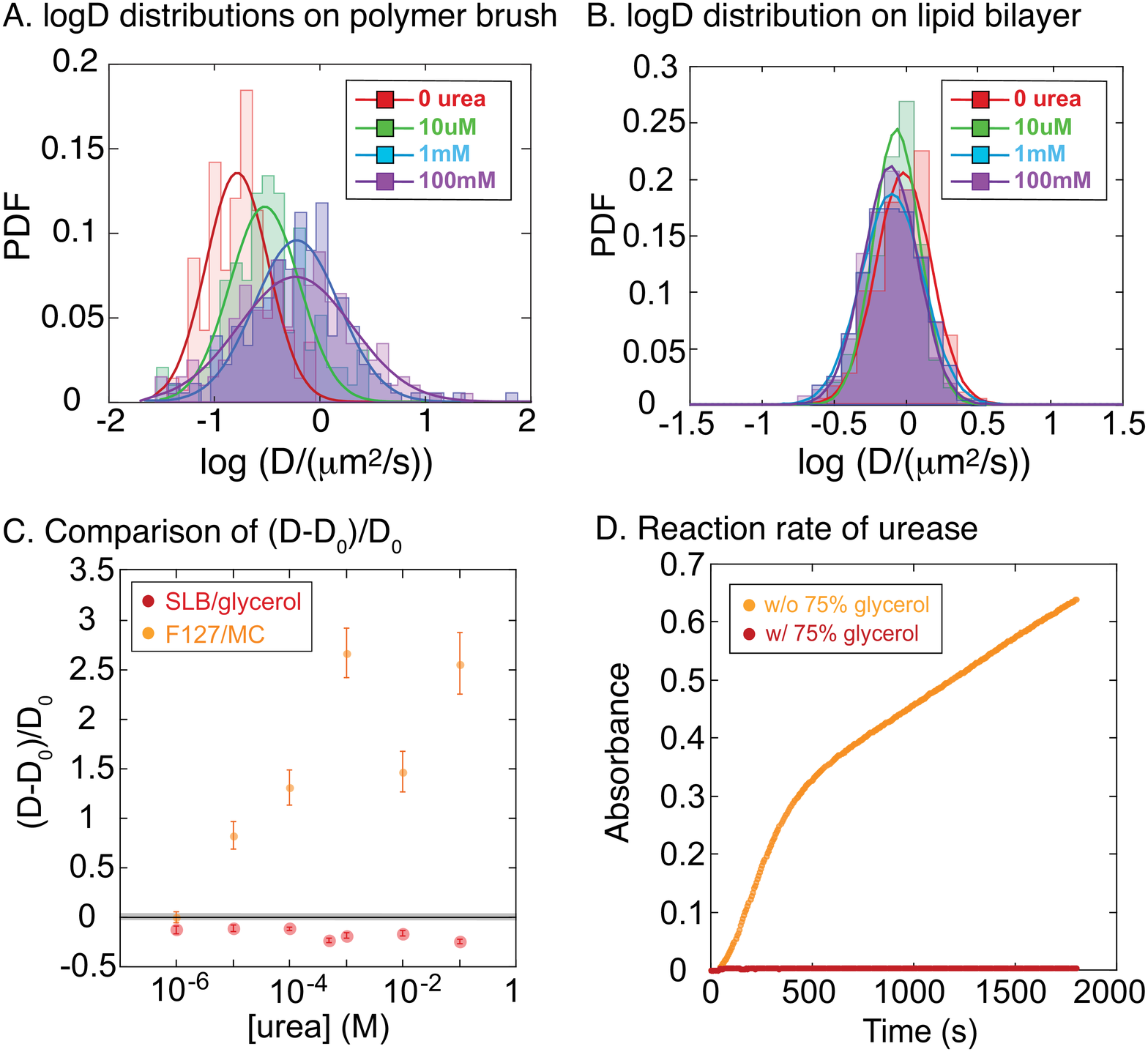}
		\caption{A) Representative probability distribution of log-transformed diffusion constants log$D$ at four different urea concentrations: $0$ (red region, $N$ = 141), 10 $\mu$M (green region, $N$ = 97), 1 mM (blue region, $N$ = 178), 100 mM (purple region, $N$ = 203), when using polymer brush chamber design. Colored lines show the Gaussian fits to the corresponding histograms. B) Representative histograms of logarithmic diffusion coefficients at different urea concentrations: $0$ (red region, $N$ = 178), 10 $\mu$M (green region, $N$ = 205), 1 mM (blue region, $N$ = 390), 100 mM (purple region, $N$ = 357) when using SLB/glycerol chamber. Colored lines show the Gaussian fits to the corresponding log$D$ histograms. C) The relative increase in $D$, $(D-D_0)/D_0$, as a function of urea concentration observed in the prior polymer brush chamber (orange dots) and the SLB/glycerol chamber (red dots), where $D_0$ is the diffusion constant when no urea is present. Error bars are determined from the standard errors of the mean of the Gaussian fits. All fit parameters are given in the supplemental information. D) Comparisons of urease-catalyzed reaction rate with (red) and without (orange) the presence of 75\% glycerol.}
	\label{fig:F127&lipid}
\end{figure}

\paragraph{Diffusion of Active Urease in the High Viscosity Regime}
We first measure the diffusion of urease at seven different urea concentrations using the SLB/glycerol chamber with 75\% glycerol ($\eta=41.3$ mPa$\cdot$s). We choose 75\% because it is the lowest glycerol percentage for MSD analysis that yields accurate diffusion coefficients. Histograms of log-transformed diffusion coefficients of urease are illustrated in Fig. \ref{fig:F127&lipid} B for four representative urea concentrations and are compared with our prior results when using F127 polymer brush chambers (Fig. \ref{fig:F127&lipid} A from Ref.\cite{Xu2019}). In the prior chamber design, urease appeared to diffuse faster with the presence of only 10 $\mu$M urea in solution. However, with the newly-designed SLB/glycerol chamber, we do not find any enhancement in urease diffusion even at the saturation concentration of urea (100 mM). To illustrate the relative increase in diffusion coefficient of urease more clearly, we plot the relative changes in $D$, $(D-D_0)/D_0$, as a function of urea concentration (Fig. \ref{fig:F127&lipid} C), where $D_0$ is the diffusion rate when no urea is present. In contrast to the $\sim$3-fold increase previously observed in the polymer brush chamber (Fig. \ref{fig:F127&lipid} C, orange dots), no relative increase is observed for urease diffusing in the SLB/glycerol chamber at any urea concentrations (Fig. \ref{fig:F127&lipid} C, red dots).

Given our ability to reproduce the enhanced diffusion of urease in the presence of urea by using the polymer brush chamber \cite{Xu2019} and by tethering urease on SLB (Fig. \ref{fig:tethered urease}), we were surprised by the lack of enhancement in the SLB/glycerol chamber. The main difference among these experiments is the high percentage (75\%) of glycerol used in the SLB/glycerol chamber. We speculate that such high amounts of glycerol might interfere with the urease activity, resulting in the failure to observe the enhanced diffusion. In order to determine if 75\% glycerol poisons the enzyme activity, we measure the activity of urease with and without the presence of 75\% glycerol using a colormetric assay. We find that the urease activity is completely inhibited by the presence of 75\% glycerol (Fig. \ref{fig:F127&lipid} D). Thus, with no catalytic activity, no matter how much substrate is present in the solution, the enzyme diffuses as it does in buffer, and no enhanced diffusion is observed. Interestingly, this result is the opposite of what was previously reported for urease activity in high glycerol in an original paper published in 1967 \cite{BLATTLER1967}. We believe that the modern techniques we employ here are better for addressing these questions than the technique used over 50 years ago. Therefore, when choosing viscous agents for enzyme diffusion experiments, extra care must be taken to make sure the enzyme activity is preserved. \\

\paragraph{Diffusion of Active Urease at Low Viscosity Regime}

\begin{figure}
	\centering
		\includegraphics[width=1\linewidth]{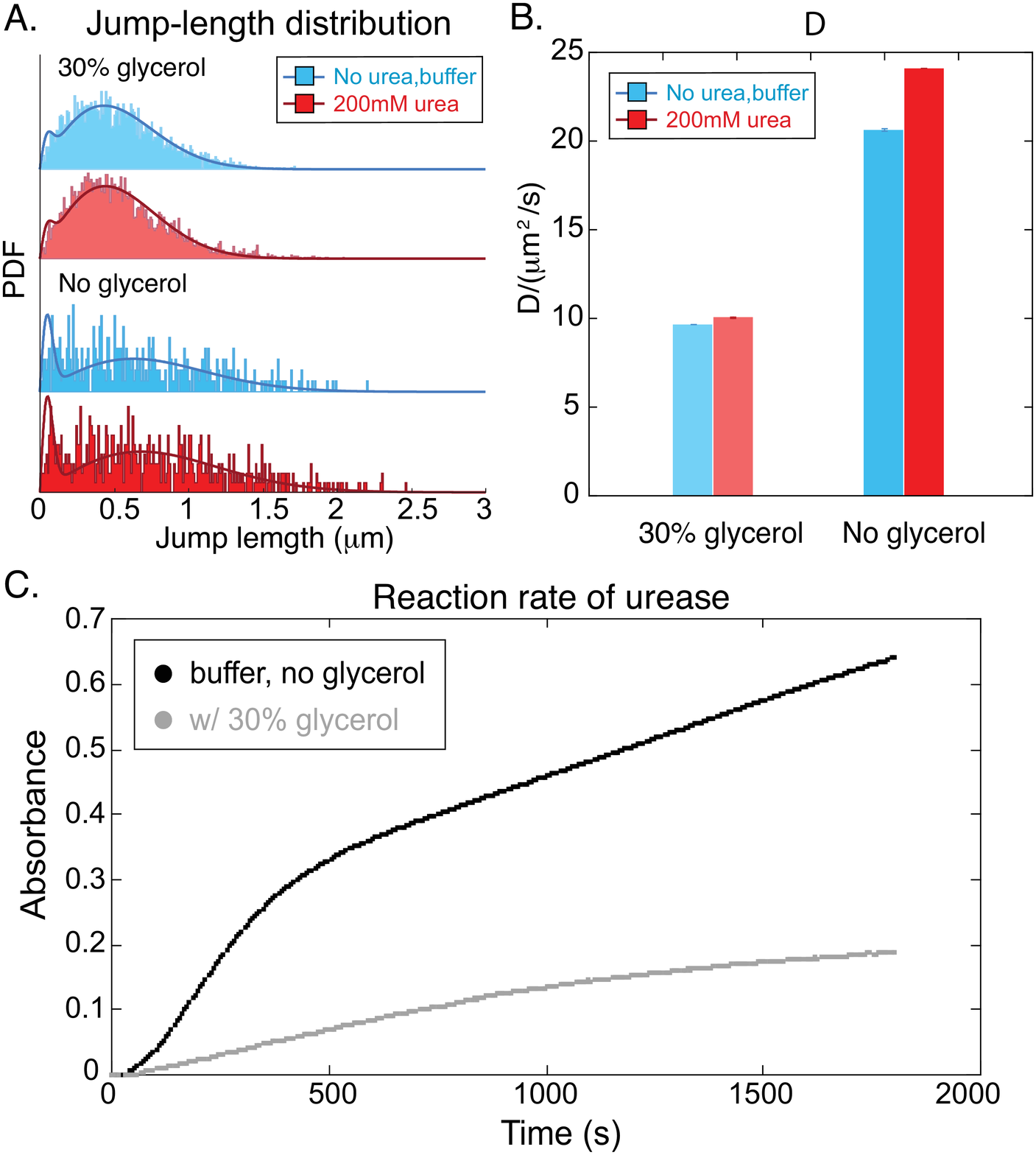}
		\caption{A) Jump-length distributions of urease at $\Delta t = 9.5$ ms with (red) and without (blue) the presence of 200 mM urea in 30\% glycerol (top) and in buffer solution (bottom). Colored line shows the corresponding kinetic model fit from Spot-On analysis for each jump-length distribution. B) Apparent diffusion coefficients reported by Spot-On analysis for urease diffusing in 30\% glycerol (left) and in buffer solution (right) with (red) and without (blue) the presence of 200 mM urea. C) Comparisons of urease-catalyzed reaction rate with (gray) and without (black) the presence of 30\% glycerol.}
	\label{fig:urease fast diffusion}
\end{figure}

We next quantify the diffusion of active urease at low viscosities using the jump-length analysis. Based on what we have found in the former section, 30\% glycerol ($\eta=2.7$ mPa$\cdot$s) appears to be an appropriate viscosity range for jump-length analysis to work. We first measure the urease diffusion in the absence and presence of 200 mM urea at 30\% glycerol. Fig. \ref{fig:urease fast diffusion} A (top) shows the jump-length distributions of urease with and without the presence of urea in our 30\% glycerol chamber. No obvious shift is observed for the urease jump-length when urea is present. After kinetic model fitting from Spot-On, we derive the diffusion coefficient for each case and plot the results in Fig. \ref{fig:urease fast diffusion} B (left). We find that for the buffer case (no urea) $D_\textrm{buffer, 30$\%$ gly}$=9.65 $\mu$m$^2$/s, while for the urea case $D_\textrm{urea, 30$\%$ gly}$=10.04 $\mu$m$^2$/s. Consistent with what has been implied by the jump-length distributions in Fig. \ref{fig:urease fast diffusion} A (top), almost no relative increase (only $\sim$4\%) in $D$ is found for urease. 

To examine the reason for the lack of enhancement, we perform the same colormetric assay for urease to check its activity in 30\% glycerol. We find that although urease still remains active under 30\% glycerol, its catalytic activity is moderately suppressed (Fig. \ref{fig:urease fast diffusion} C). The enzymatic catalysis rate is not as fast as before. This implies that even a slight amount of change on enzyme activity might result in the failure to observe the enhanced enzyme diffusion. 

To avoid adding glycerol, we make the same diffusion measurements for urease in buffer solution (no glycerol, $\eta=1$ mPa$\cdot$s). Fig. \ref{fig:urease fast diffusion} A (bottom) and Fig. \ref{fig:urease fast diffusion} B (right) show the jump-length distributions and the apparent diffusion coefficients reported by Spot-On analysis, respectively. With no glycerol present, a slight increase $\sim$17\% in $D$ is observed for urease diffusing in urea solution ($D_\textrm{buffer, no gly}$=20.65 $\mu$m$^2$/s, $D_\textrm{urea, no gly}$=24.09 $\mu$m$^2$/s). This 17\% increase is much lower than the 3-fold enhancement that we have observed in our prior polymer brush design \cite{Xu2019} or the tethered urease experiments (Fig. \ref{fig:tethered urease}). Interestingly, this slight enhancement is similar to the increase reported in the prior studies of enhanced urease diffusion using FCS measurements \cite{Muddana2010, Riedel2015, Jee2019}.

We surmise that this underestimation of relative increase in $D$ is likely due to the inaccurate tracking of very fast-moving particles, which is limited by the spatial and temporal scales set in our TIRF microscope. Several facts imply that this may be the case: 1) the apparent diffusion coefficients derived from the jump-length analysis are much lower than the expected value estimated from the Stokes-Einstein equation ($D_\textrm{expected}$=31.38 $\mu$m$^2$/s); 2) the noisiness of the jump-length distributions in the absence of glycerol (Fig. \ref{fig:urease fast diffusion} A (bottom)) indicates that fewer trajectories are analyzed compared to the 30\% glycerol scenario (Fig. \ref{fig:urease fast diffusion} A (top)). Few data points could result in inappropriate model fitting and an inaccurate $D$. Therefore, at extremely low viscosities, even jump-length analysis may not be applicable. In the first Spot-On analysis paper, the authors only tested $D$ within the range of $0.5\sim14.5$ $\mu$m$^2$/s \cite{Hansen2018}, implying that this method might not be able to capture faster-diffusing particles. Despite that, follow-up papers have used the Spot-On analysis to measure diffusion rates as fast as $\sim$50 $\mu$m$^2$/s, giving us confidence to try this method on our cases \cite{Chen2020}. 

In summary, we examine the diffusion of active urease with and without urea using SLB/glycerol chamber at two viscosity regimes. These experiments do not reproduce the enhanced diffusion of urease due to the the presence of glycerol poisoning enzyme activity. To avoid adding glycerol, we make the same urease diffusion measurements in buffer solution (no glycerol), yet we find that our analysis methods are unable to achieve accurate tracking and adequate data analysis at such low viscosity. Our results imply that a moderate slowing down of diffusion is needed for accurate SPT of enhanced enzyme diffusion, but care must be taken when choosing viscosity agents to preserve enzyme activity.

\section{conclusion}

The aim of this work is to compare different experimental designs and different data analysis methods for single-molecule imaging of enhanced enzyme diffusion experiments. We find that the 2D confinement of urease to a fluid lipid bilayer conserves enhanced enzyme diffusion, recapitulating the 3-fold enhancement of urease at the saturation concentration of urea, as we previously reported in Ref.\cite{Xu2019}. To minimize the chamber complexity, we design a new chamber with a well-characterized SLB coated surface and a small viscous molecule, glycerol, to replace the bulky polymers used in the prior F127 polymer brush design. We confirm the efficacy of the newly-designed SLB/glycerol chamber by measuring the diffusion of different-size particles in different viscosity solutions. We find that particles diffuse as the Stokes-Einstein equation predicts: their diffusion coefficients scale inversely with solvent viscosity $\eta$ and particle size $R$. 

We also compare two data analysis methods for SPT: the MSD analysis and the jump-length analysis. We find that MSD analysis is appropriate for analyzing slowly diffusing species, when high solvent viscosity or large particle size are preferable. While analyzing fast diffusion, MSD analysis under-counts the population of fast-moving particles, leading to an underestimation of the actual diffusion coefficient. Jump-length analysis seems to be applicable for a wider range, from very slow diffusion to relatively fast motion. However, we also notice that for jump-length analysis the diffusion coefficients reported depend heavily on the parameters, especially one of the input parameters in MTT tracking algorithm, $D_\textrm{max}$. Thus, care must be taken when choosing parameters and a dataset of all key parameters used for the analysis should be reported specifically to allow for reproducibility and transparency when using this method. 

We next examine how urease diffuses in the SLB/glycerol chamber with and without the presence of urea. We measure the diffusion of urease at two viscosity regimes: the high viscosity regime with 75\% glycerol and the low viscosity regime with 30\% glycerol. However, no enhanced diffusion is observed for urease at either viscosity due to the inactivation of urease by glycerol. When we perform the same urease diffusion experiments in buffer solution without glycerol, the enzymes diffuse too fast in buffer to allow for accurate diffusion measurements. 

Taken together, we find that the previously employed F127 polymer brush chamber seem to be excellent at slowing down enzyme motility without inhibiting its activity. The tethered enzyme experimental design demonstrated in this work is also a viable strategy. Overall, an optimized experimental design, as well as a more intuitive, less parameter-dependent data analysis approach, are still needed for future investigations of enhanced enzyme diffusion.

\begin{acknowledgments}
We want to thank Arnauld Sergé and Anders Sejr Hansen for their helpful suggestions about the parameter choosing for MTT and Spot-On analysis. MX was partially supported by NSF MRSEC DMR-1420382 to Seth Fraden (Brandeis University). All investigation partially supported by NSF DMR-2004417 to JLR, NSF DMR-2004400 to WBR, and NSF DMR-2004566 to WWA. MX and JLR were also supported in part from start-up funds from Syracuse University.
\end{acknowledgments}

\bibliographystyle{apsrev4-1} 
\bibliography{citation.bib} 


\end{document}


\title{Supplemental Information: Comparison of different approaches to single-molecule imaging of enhanced enzyme diffusion}

\author{Mengqi Xu}
\affiliation{Department of Physics, Syracuse University,  Syracuse, NY 13244}
\author{W. Benjamin Rogers}
\affiliation{Martin A. Fisher School of Physics, Brandeis University, Waltham, MA 02453 USA}
\author{Wylie W. Ahmed}
\affiliation{Department of Physics, California State University, Fullerton, CA 92831}
\author{Jennifer L. Ross}
\affiliation{Department of Physics, Syracuse University,  Syracuse, NY 13244}

\maketitle
\textbf{Gaussian fit:}

\begin{equation}
   PDF = A\times \exp\big[-\frac{(log(D)-<log(D)>)^2}{2 \sigma^2}\big] , \label{Gaussian}
    \label{eqn:Gaussian}
\end{equation} 

We fit the histograms of log$D$ with equation (1). The mean of each Gaussian fit is transformed back to the typical diffusion units, acting as the apparent diffusion coefficient for each case. Error bars are obtained from the standard error of the each Gaussian fit. The top of the error bar is determined by adding the mean by the standard error to determine the right-most edge of the Gaussian width and then taking that as the power of 10 to transform it back to $D$. The bottom of the error bar is determined by the same way except subtracting the standard error. The fit parameters for each experiment are given in the table 1-4. ${\chi}^2$ denotes the Chi-Square goodness of the fit test.\\ \\

\begin{table}[h]
\label{tb1}
\centering
\begin{tabular}{@{}lcccccccccc@{}}\toprule
[urea]          && $N$     && $A$                 &   & $<$log(D/($\mu$m$^2$/s))$>$     &  & $\sigma$     &  & ${\chi}^2$     \\
\cline{1-11}
buffer          && 178     && 0.047 $\pm$ 0.002   &   & -1.084 $\pm$ 0.050   &  & 1.176 $\pm$ 0.071    &  & 0.0035         \\
200 mM        && 484     && 0.053 $\pm$ 0.002   &   & -0.627 $\pm$ 0.040   &  & 1.371 $\pm$ 0.077    &  & 0.0034         \\
\cline{1-11}
\end{tabular}
\caption{Fit parameters to Gaussian fit equation (1) for log transformed diffusion data under each urea concentration shown in Figure 2A, B. $N$ denotes the number of trajectories in each distribution histogram and is used to calculate the standard error of each Gaussian fit.} 
\label{tab: fit of D fig2}
\end{table}

\clearpage

\begin{table}[h]
\label{tb1}
\centering
\begin{tabular}{@{}lcccccccccc@{}}\toprule
glycerol\%          && $N$    && $A$               &   & $<$log(D/($\mu$m$^2$/s))$>$       &  & $\sigma$           &  & ${\chi}^2$     \\
\cline{1-11}
30\%         && 10      && 0.359 $\pm$ 0.031   &   & 0.061 $\pm$ 0.008   &  & 0.082 $\pm$ 0.008    &  & 0.0331         \\
50\%         && 420     && 0.168 $\pm$ 0.004   &   & 0.055 $\pm$ 0.007   &  & 0.223 $\pm$ 0.007    &  & 0.0019         \\
60\%         && 109     && 0.116 $\pm$ 0.006   &   & 0.202 $\pm$ 0.020   &  & 0.339 $\pm$ 0.020    &  & 0.0052         \\
70\%         && 313     && 0.131 $\pm$ 0.005   &   & 0.092 $\pm$ 0.013   &  & 0.303 $\pm$ 0.013    &  & 0.0033         \\
75\%         && 97      && 0.196 $\pm$ 0.006   &   & 0.029 $\pm$ 0.008   &  & 0.198 $\pm$ 0.008    &  & 0.0037         \\
80\%         && 676     && 0.176 $\pm$ 0.004   &   & -0.162 $\pm$ 0.006   &  & 0.213 $\pm$ 0.006    &  & 0.0018         \\
85\%         && 736     && 0.143 $\pm$ 0.002   &   & -0.341 $\pm$ 0.005   &  & 0.269 $\pm$ 0.005    &  & 0.0006         \\
90\%         && 213     && 0.122 $\pm$ 0.004   &   & -0.533 $\pm$ 0.013   &  & 0.313 $\pm$ 0.013    &  & 0.0026         \\
\cline{1-11}
\end{tabular}
\caption{Fit parameters to Gaussian fit equation (1) for log transformed diffusion data under each urea concentration shown in Figure 3A, C. $N$ denotes the number of trajectories in each distribution histogram and is used to calculate the standard error of each Gaussian fit.} 
\label{tab: fit of D fig3}
\end{table}

\begin{table}[h]
\label{tb1}
\centering
\begin{tabular}{@{}lcccccccccc@{}}\toprule
Particle         && $N$     && $A$                 &   & $<$log(D/($\mu$m$^2$/s))$>$       &  & $\sigma$           &  & ${\chi}^2$     \\
\cline{1-11}
GFP              && 49      && 0.135 $\pm$ 0.013   &   & 0.073 $\pm$ 0.030   &  & 0.268 $\pm$ 0.030    &  & 0.0218         \\
aldolase         && 97      && 0.196 $\pm$ 0.006   &   & 0.029 $\pm$ 0.007   &  & 0.198 $\pm$ 0.007    &  & 0.0037         \\
urease           && 113     && 0.180 $\pm$ 0.008   &   & 0.097 $\pm$ 0.016   &  & 0.220 $\pm$ 0.016    &  & 0.0129         \\
plastic sphere   && 46      && 0.147 $\pm$ 0.011   &   & -1.223 $\pm$ 0.017   &  & 0.278 $\pm$ 0.017    &  & 0.0082         \\
\cline{1-11}
\end{tabular}
\caption{Fit parameters to Gaussian fit equation (1) for log transformed diffusion data under each urea concentration shown in Figure 4A, C. $N$ denotes the number of trajectories in each distribution histogram and is used to calculate the standard error of each Gaussian fit.} 
\label{tab: fit of D fig4}
\end{table}

\begin{table}[h]
\label{tb1}
\centering
\begin{tabular}{@{}lcccccccccc@{}}\toprule
[urea]          && $N$    && $A$               &   & $<$log(D/($\mu$m$^2$/s))$>$       &  & $\sigma$           &  & ${\chi}^2$     \\
\cline{1-11}
buffer          && 178     && 0.205 $\pm$ 0.009   &   & -0.011 $\pm$ 0.009   &  & 0.193 $\pm$ 0.009    &  & 0.0065         \\
1 $\mu$M        && 188     && 0.200 $\pm$ 0.004   &   & -0.074 $\pm$ 0.005   &  & 0.201 $\pm$ 0.005    &  & 0.0017         \\
10 $\mu$M       && 205     && 0.244 $\pm$ 0.009   &   & -0.060 $\pm$ 0.007   &  & 0.160 $\pm$ 0.007    &  & 0.0059         \\
100 $\mu$M      && 701     && 0.207 $\pm$ 0.004   &   & -0.055 $\pm$ 0.004   &  & 0.191 $\pm$ 0.004    &  & 0.0014         \\
500 $\mu$M      && 456     && 0.208 $\pm$ 0.003   &   & -0.113 $\pm$ 0.003   &  & 0.190 $\pm$ 0.003    &  & 0.0006         \\
1 mM            && 390     && 0.187 $\pm$ 0.002   &   & -0.092 $\pm$ 0.003   &  & 0.215 $\pm$ 0.003    &  & 0.0004         \\
10 mM           && 383     && 0.199 $\pm$ 0.007   &   & -0.078 $\pm$ 0.008   &  & 0.201 $\pm$ 0.008    &  & 0.0043         \\
100 mM          && 357     && 0.211 $\pm$ 0.003   &   & -0.101 $\pm$ 0.004   &  & 0.189 $\pm$ 0.004    &  & 0.0010         \\
\cline{1-11}
\end{tabular}
\caption{Fit parameters to Gaussian fit equation (1) for log transformed diffusion data under each urea concentration shown in Figure 5B, C. $N$ denotes the number of trajectories in each distribution histogram and is used to calculate the standard error of each Gaussian fit.} 
\label{tab: fit of D fig5}
\end{table}

\clearpage

\textbf{Parameter settings used for MTT algorithm:}

\begin{table}[ht]
\centering
\begin{tabular}{l c c c}
\hline\hline
Experiments & ExposureTime & MaxExpectedD ($D_\textrm{max}$) & NumGapsAllowed \\ [0.5ex]
\hline
aldolase in 90\% glycerol     & 60      & 1.4   &    3 \\
aldolase in 85\% glycerol     & 60      & 2     &    3 \\
aldolase in 80\% glycerol     & 60      & 2     &    3  \\
aldolase in 75\% glycerol     & 60      & 5     &    3  \\
aldolase in 70\% glycerol     & 60      & 7     &    3  \\
aldolase in 60\% glycerol     & 60      & 8     &    3  \\
aldolase in 50\% glycerol     & 60      & 20    &    3  \\
aldolase in 30\% glycerol     & 60      & 25    &    3  \\
GFP in 75\% glycerol          & 60      & 10    &    3  \\
urease in 75\% glycerol       & 60      & 3     &    3  \\
plastic spheres in 75\% glycerol             & 60   & 1   &    3       \\
urease in 30\% glycerol with no urea         & 9.5  & 30  &    2       \\
urease in 30\% glycerol with 200mM urea      & 9.5  & 40  &    2       \\
urease in buffer solution with no urea       & 9.5  & 50  &    2       \\
urease in buffer solution with 200mM urea    & 9.5  & 70  &    2       \\ [1ex]
\hline
\end{tabular}
\caption{Parameter settings used in MTT algorithm for each experimental case, with LocalizationError = -6.25, EmissionWavelength = 647, NumDeflationLoops = 0 for all cases.} 
\label{table:parameter for MTT}
\end{table}